\documentclass[
 reprint,
superscriptaddress,
 amsmath,amssymb,
 aps,
floatfix,
,colorlinks=true,linkcolor=black,citecolor=black,pdfencoding=auto]{revtex4-1}

\usepackage{braket}
\usepackage{dsfont}
\usepackage{graphicx}
\usepackage[dvipsnames]{xcolor}
\usepackage{color}
\usepackage{comment}
\usepackage{appendix}
\usepackage[colorlinks=true,
            linkcolor=black,
            citecolor=blue,
            pdfencoding=auto]{hyperref}
\usepackage{blindtext}

\renewcommand{\bm}[1]{{\bf  #1}}

\begin{document}

\title{Unified Theory of the Anomalous and Topological Hall Effects with Phase Space Berry Curvatures}

 \author{Nishchhal Verma}
\thanks{These authors contributed equally}
\affiliation{Department of Physics,
The Ohio State University,
Columbus, OH 43210, USA}
\author{Zachariah Addison}
    \thanks{These authors contributed equally}
\affiliation{Department of Physics,
The Ohio State University,
Columbus, OH 43210, USA}
\author{Mohit Randeria}
\affiliation{Department of Physics,
The Ohio State University,
Columbus, OH 43210, USA}

\begin{abstract}
Hall experiments in chiral magnets are often analyzed as the sum of an anomalous Hall effect, 
dominated by momentum-space Berry curvature, and a topological Hall effect, arising from the real-space Berry curvature in the presence
of skyrmions, in addition to the ordinary Hall resistivity. 
This raises the questions of how one can incorporate, on an equal footing, the effects of the anomalous velocity and 
the real space winding of the magnetization, and when such a decomposition of the resistivity is justified.
We provide definitive answers to these questions by including the effects of all phase-space Berry curvatures in a semi-classical approach and by
solving the Boltzmann equation in a weak spin-orbit coupling regime when the magnetization texture varies slowly on 
the scale of the mean free path.
We show that the Hall resistivity is then just the sum of the anomalous and topological contributions, with negligible corrections from 
Berry curvature-independent and mixed curvature terms. We also use an exact Kubo formalism to numerically investigate the 
opposite limit of infinite mean path, and show that the results are similar to the semi-classical results.
\end{abstract}

\maketitle
The Hall effect in magnetic materials has a long history \cite{Hall1881}.
One might be tempted to think the primary explanation for the effect is that the magnetization ${\bf M}$ can exert a Lorentz force on the materials electrons, 
however, this effect is negligible for the non-relativistic charge carriers in metals \cite{WannierPR1947}.
Karplus and Luttinger \cite{KarplusPR1954} correctly identified the importance of spin-orbit coupling (SOC) 
in the anomalous Hall effect (AHE); their analysis is now best understood in terms of 
the momentum-space Berry curvature of the electron \cite{NagaosaJPSJ2006,NagaosaRMP2010,Xiao2010}.
Scattering in the presence of SOC also makes extrinsic contributions \cite{Smit1955,Berger1964} to the AHE, but the 
anomalous velocity that arises from Berry curvature effects is an intrinsic effect that dominates in many experiments \cite{NagaosaRMP2010,YaoPRL2004}, and will be our focus here.

Understanding the Hall effect with spatially varying magnetic textures poses further challenges.
In addition to the AHE, experiments see a topological Hall effect (THE)
in a variety of chiral magnetic materials that harbor skyrmions, including B20 crystals \cite{LeePRL2009,NeubauerPRL2009,KanazawaPRL2011} and thin films \cite{LiPRL2013,GallagherPRL2017,AhmedPRM2018}, 
and heavy metal/magnetic insulator bilayers \cite{AhmedACS2019,ShaoNE2019}.
Skyrmions give rise to an emergent magnetic field that derives from the real-space Berry curvature,
resulting in a THE proportional to their topological charge density $n_{\text{sk}}$ \cite{YePRL1999,Tatara2002,Bruno2004,OnodaJSPJ2004,NagaosaRSTA2012,NagaosaNT2013,HamamotoPRB2015,NakazawaJSPJ2018, Ishizuka2018}.

Theories of the anomalous and topological Hall effect have for the most part been distinct and, 
despite important recent progress \cite{Kim2013,Akosa2018,LuxNature2018,Akosa2019,BatistaPRB2020,LuxPRL2020,BouazizPRL2021}
on electrons with SOC interacting with skyrmions, a single theory that incorporates both real and momentum space 
Berry curvature effects on an equal footing to calculate electronic transport has remained elusive.
The experiments \cite{LeePRL2009,NeubauerPRL2009,KanazawaPRL2011,LiPRL2013,GallagherPRL2017,AhmedPRM2018,AhmedACS2019,ShaoNE2019},
on the other hand, are routinely interpreted as a sum of an anomalous and topological Hall resistivity, in addition to the ordinary Hall effect proportional to the magnetic field.

\begin{figure*}
\includegraphics[width=\textwidth]{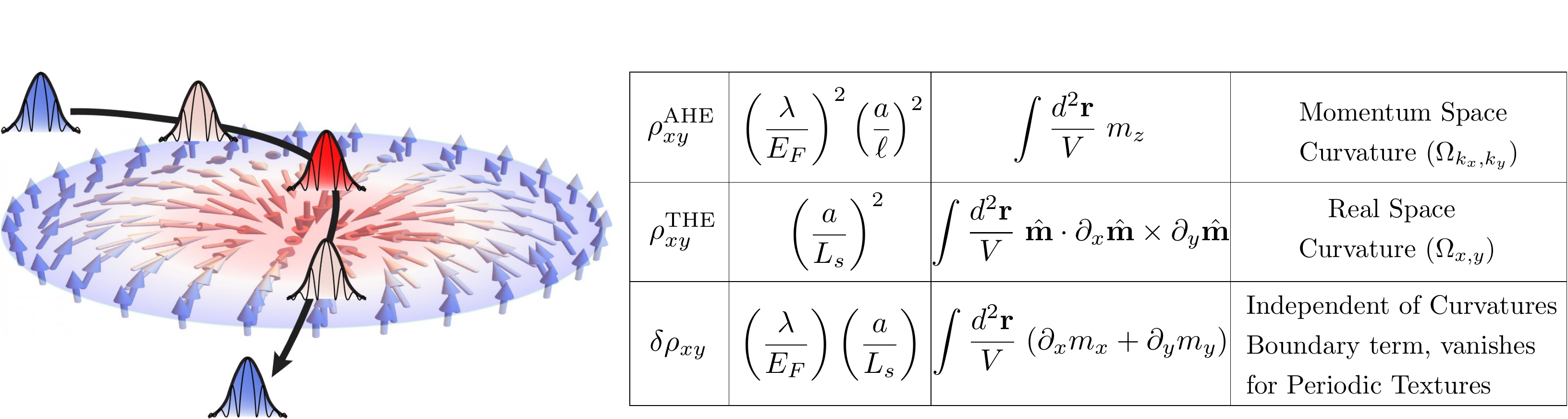}
\caption{
The semi-classical wave-packet follows the texture and is influenced by real-space Berry curvature arising from the presence of skyrmions, 
in addition to the anomalous velocity that it acquires from an external electric field and momentum space Berry curvature. 
Our results are obtained in the regime where spin texture length scale $L_s \gg$ mean free path $\ell \gg a$, the lattice spacing, 
and weak spin-orbit coupling $\lambda \ll E_F$, the Fermi energy.
The table summarizes the three contributions to $\rho_{xy}$, their scaling with these parameters, 
their dependence on the magnetic texture $\hat{\bf m}({\bf r})$, and their relation to Berry curvatures.
Mixed momentum- and real-space curvatures contribute to the Hall resistivity at higher order in $(\lambda/E_F)$ and $(a/L_s)$.}
 \label{tab-scaling}
\end{figure*}

In this paper, we demonstrate within a semi-classical theory that the Hall response (excluding the ordinary Hall effect) is
just the sum of two terms, the AHE and the THE.
The semi-classical approach  \cite{Xiao2010, FreimuthPRB2013} is a natural avenue to study the effects of all phase space Berry curvatures, $\bm{r}$-space, $\bm{k}$-space, and mixed, on an equal footing in the regime where the length scale $L_s$ on which the 
spin texture varies and the mean free path $\ell$ from impurity scattering are both much larger than the microscopic
scales of the average inter-particle spacing $k_F^{-1}$ or the lattice spacing $a$. We are also interested in the regime of weak
SOC $\lambda \ll E_F$, the Fermi energy. 

To determine the Hall resistivity we solve the Boltzmann equation to linear order in the electric field in the presence of all phase-space curvatures and real and momentum space derivatives of the semi-classical energy eigenvalues.  
Systematically classifying the resulting array of terms in powers of the small parameters $\lambda/E_F$ and $\ell/L_s$,
and extracting the leading contributions in the regime $L_s \gg \ell \gg k_F^{-1} \simeq a$ and $\lambda \ll E_F$,
we find that 
\begin{equation}
\rho_{xy} = \rho_{xy}^{\text{AHE}} + \rho_{xy}^{\text{THE}} + \delta\rho_{xy}.
\end{equation}

Our results are summarized in the table in Fig.~1, where we show how each term depends (i) on the small parameters 
that control our calculation, (ii) on the spatially varying magnetization ${\bf M} = M_s\, \hat {\bf m}({\bf r})$, and 
(iii) on the Berry curvatures. While the first two terms represent the AHE and the THE respectively, the correction
term $\delta\rho_{xy}$ is a curvature-independent boundary contribution proportional to the vorticity of the local electronic velocity field. It vanishes when the spin texture is
periodic, e.g., a skyrmion crystal, and is negligible for a disordered skyrmion array in the thermodynamic limit.
We show that the mixed curvatures contribute to the Hall resistivity at higher order in the 
small parameters $(\lambda/E_F)$ and $(a/L_s)$ than the terms shown in Fig.~1.
 
Finally, we also present results using the Kubo formula in the opposite regime 
where $\ell \gg L_s \gtrsim k_F^{-1} \simeq a$.
We focus on a disorder free system with $\ell = \infty$, use exact diagonalization 
in the magnetic unit cell of a skyrmion crystal, and compute the total Hall conductivity using the 
TKNN formula \cite{TKNN1982} in the magnetic Brillouin zone, which includes the effects of both 
the anomalous velocity and of the skyrmion topological charge density.
We show how the semi-classical results allow us to qualitatively understand 
all of the non-trivial parameter dependencies of the Hall response including the dependence on the density, the SOC, and the 
exchange coupling between the charge carriers and the spin.

\medskip

{\bf Model:}
We analyze a minimal Hamiltonian for studying the confluence of anomalous and topological Hall effects.
It can arise either from a ``s-d model" of itinerant electrons interacting with local moments in a metallic magnet with Rashba SOC, or
alternatively, it can be used to model the conduction electrons in a metal proximate to a magnetic insulator 
where broken inversion symmetry at the interface induces a Rashba SOC. 

We consider a 2D Hamiltonian
\begin{equation}
\widehat{\mathcal{H}}  = \dfrac{ \widehat{{\bf p}}^2 }{2m} + \dfrac{a\lambda}{\hbar} \left( \widehat{{\bf p}} \times \hat{{\bf z}} \right) \cdot \boldsymbol\sigma  
- J \;\hat{{\bf m}}( \widehat{{\bf r}}) \cdot \boldsymbol\sigma + \widehat{\mathcal{H}}_\text{imp} \label{eq:ham-gen}
\end{equation}
which describes itinerant electrons of mass $m$ and Rashba SOC $\lambda$ whose spin $\boldsymbol\sigma$ is
coupled to a magnetic texture ${\bf M} = M_s\, \hat{{\bf m}}({\bf r})$ via an exchange interaction $J$. 
Elastic scattering of electrons off a disorder potential is described by 
$\widehat{\mathcal{H}}_\text{imp}$ and leads to a mean free path $\ell \gg k_F^{-1}$.
The small hats denote unit vectors and the wide hats denote quantum mechanical operators.
Based on the separation of time-scales associated with the itinerant electrons and the dynamics of spins in the texture, 
we assume that the texture is static. 
The model has three energy scales: the Fermi energy $E_F$, SOC $\lambda$, and exchange coupling $J$, and three length scales: the inter-particle spacing $k_F^{-1}$ ($\approx a$, the lattice spacing), the mean-free path $\ell$, and the length scale $L_s$ associated to the spatial variations of the magnetic texture.
We will focus on the weak SOC regime $\lambda \ll J, E_F$, relevant for experiments.

\medskip

{\bf Semi-classical Equations of Motion:}
Let us focus on the semi-classical regime $L_s \gg k_F^{-1}$.
To analyze the dynamics of electron wave packets in phase space $\boldsymbol\xi = (x,y,k_x, k_y)$, 
we follow the standard prescription \cite{Xiao2010} to construct the semi-classical Hamiltonian 
\begin{equation}
\mathcal{H}(\boldsymbol\xi) = \dfrac{ \hbar^2 {\bf k}^2 }{2m} + {\bf d}(\boldsymbol\xi) \cdot {\boldsymbol\sigma}
\label{ham-sc-1}
\end{equation}
where ${\bf d}(\boldsymbol\xi) = a\lambda ( {\bf k} \times \hat{{\bf z}} ) - J \hat{{\bf m}}({\bf r})$ captures the 
quantum mechanical nature of the spin. The semi-classical eigenenergies are 
$\mathcal{E}_\pm(\boldsymbol\xi) =\hbar^2 {\bf k}^2/2m \pm | {\bf d} (\boldsymbol\xi) |$.
The corresponding wavefunctions posses non-trivial phase space geometry encoded in the Berry curvatures
\begin{eqnarray}
\Omega^\pm_{ \alpha, \beta }(\boldsymbol\xi) &=& \pm \dfrac{1}{2} \hat{ {\bf d} }(\boldsymbol\xi) \cdot \left( \partial_{\alpha } \hat{\bf d}(\boldsymbol\xi) \times \partial_{\beta } \hat{{\bf d}}(\boldsymbol\xi) \right)
\end{eqnarray}
each corresponding to one of the six orthogonal planes in the 4D phase space spanned by $\boldsymbol\xi$.
The dynamics of the semi-classical theory describe intra-band processes such that each electronic band may be treated independently,
and we will suppress the band index unless necessary.

The curvatures modify the equations of motion as well as the invariant measure in phase space.
To simplify notation, we introduce a $4 \times 4$ matrix,
\begin{equation}
[\Gamma(\boldsymbol\xi)]_{\alpha,\beta} = \Omega_{ \alpha, \beta}(\boldsymbol\xi)-[ i\sigma_y \otimes \mathds{1} ]_{\alpha,\beta}
\end{equation}
to write the equations of motion
\begin{equation}
\dot{\xi}_\alpha (\boldsymbol\xi) = [\Gamma^{-1}(\boldsymbol\xi)]_{\alpha\beta} \; \big(\partial_{\beta} \widetilde{\mathcal{E}}(\boldsymbol\xi) 
+ e E \;\delta_{ \beta, y}\big)/\hbar \label{eq:eom-Gamma}
\end{equation}
where $E$ is the external electric field along the $\hat{{\bf y}}$ direction and the electron charge is $(-e)$.  
Here $\widetilde{\mathcal{E}}(\boldsymbol\xi) \simeq \mathcal{E}(\boldsymbol\xi)$ up to corrections of order 
$(\lambda/E_F)(a/L_s)$ that can be ignored in the regime of interest.
Our compact notation hides all the familiar terms, including the anomalous velocity, inside $\Gamma^{-1}$;
see appendix \ref{app:sc-equations} for more details

The combination of a spatially varying magnetic texture and SOC leads to finite real-space, momentum-space and mixed real-momentum space curvatures.
The electrons acquire an anomalous velocity proportional to the momentum-space Berry curvature $\Omega_{k_x,k_y}$, an ``anomalous force'' proportional to the real-space Berry curvature $\Omega_{x,y}$ and corrections to the group velocity and generalized force proportional to the
mixed real-momentum-space Berry curvatures.

Crucially, in addition to the equations of motion, the curvatures also modify the volume element that remains invariant under phase-space flows.
Thus to satisfy Liouville's theorem, one must use the integration measure  \cite{Xiao2010,Addison2022arxiv}
$dV_{\boldsymbol \xi} =\sqrt{\det[\Gamma(\boldsymbol \xi )]} \,d^4\boldsymbol\xi/(2\pi)^2V$, where $V$ is the volume of the system.
We note that in the presence of an external magnetic field $B_z\hat{\bm{z}}$, $\sqrt{\det[\Gamma(\boldsymbol \xi )]}$ reduces to the 
well-known factor of $(1 + e\Omega_{k_x,k_y}B_z/\hbar)$ when only the momentum-space curvature is present, 
however, we will need the more general result here.
\medskip

{\bf Hall Conductivity:}
With electric field applied along $\hat{y}$, we must calculate the transverse current along $\hat{x}$:
\begin{equation}
  j_x = -e \int dV_{\boldsymbol\xi} \; \dot{x}(\boldsymbol\xi) \; f(\boldsymbol\xi) \label{eq:curr}
\end{equation}
where $f(\boldsymbol\xi)$ is the electronic distribution function.
The distribution function reduces to the equilibrium Fermi-Dirac function $f^0[\mathcal{E}(\boldsymbol\xi)]$ in the absence of the external electric field.
The goal is to find contributions that are linear order in $E$ to calculate the electric conductivity.
 
The anomalous Hall contribution to the current derives from the intrinsic anomalous velocity and couples to the equilibrium distribution function $f^0[\mathcal{E}(\bm{\xi})]$. We isolate the terms in $\dot{x}$ linear in $E$ to find
\begin{equation}
\sigma^{\text{AHE}}_{xy} =- \dfrac{e^2}{\hbar} \sum_{l=\pm} \int \dfrac{d^2 {\bf r} \;d^2{\bf k} }{(2\pi)^2V} \;\Omega^l_{ k_x, k_y }(\boldsymbol\xi)  \; f^0_l[ \mathcal{E}_l(\boldsymbol\xi) ] \label{eq:sigma-A}
\end{equation}
where $l=\pm$ indexes the two bands.  
We emphasize that $ \sqrt{\det[\Gamma(\boldsymbol \xi )]}$ in the measure 
exactly cancels the determinant factor in $\Gamma^{-1}(\boldsymbol\xi)$ so that the final answer depends only on the momentum-space Berry curvature.
We further expand $\Omega_{k_x, k_y}(\boldsymbol\xi)$ to lowest order in $\lambda/J$ to find 
\begin{align}
\sigma_{xy}^\text{AHE} & \approx -\dfrac{e^2 a^2}{2\hbar}\,\overline{m}_z \left({\lambda}/{J}\right)^2  \sum_{l=\pm}\, l\, n_l
\label{eq-ahe}
\end{align}
where $\overline{m}_z=\int d^2\bm{r} \; \hat{\bm{m}}_z(\bm{r})/V$ is the average out-of-plane magnetization and the band-resolved density
$n_l= \int d^2\bm{k}\; f^0[\mathcal{E}_l(\bm{k})]/(2\pi)^2$ with $\mathcal{E}_l(\bm{k})=\mathcal{E}_l({\boldsymbol \xi};\lambda=0)$.  

The corresponding resistivity is found from the conductivity via 
$\rho_{xy}= -\sigma_{xy}/(\sigma^2_{xx} + \sigma^2_{xy})$ where $\sigma_{xy}\ll\sigma_{xx}=(e^2/h) k_F\ell$.  
This relationship will be used to convert conductivities to resistivities for each contribution to the Hall effect.  
For the AHE this leads to the scaling relation $\rho_{xy}^{\text{AHE}}\sim (\lambda/E_F)^2(a/\ell)^2$.

\begin{figure*}
\includegraphics[width=0.95\textwidth]{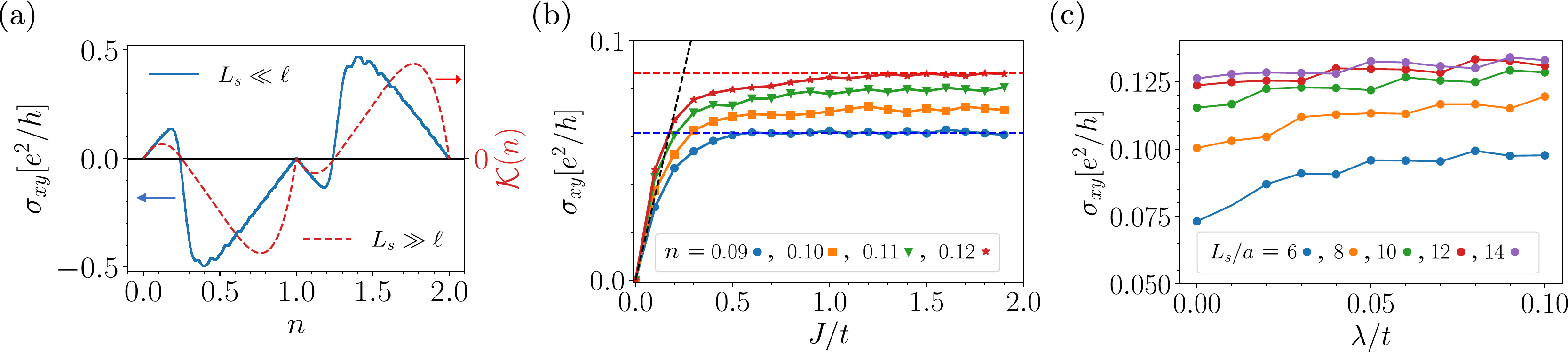}
\caption{The Hall conductivity calculated from the Kubo formula. (a) The blue curve is the THE calculated at $\lambda = 0$ with $J/t = 10$.
The red curve shows $\mathcal{K}(n) = \mathcal{K}_+ + \mathcal{K}_-$ [eq.~\eqref{eq:def-Kmu}]
which describes how the band strucure controls the semi-classical THE result~\eqref{eq:sigma-T-sc}.
Despite their different regimes of validity,
both the Kubo $\sigma_{xy}$ and $\mathcal{K}(\mu)$ have the same sign for all densities and vanish at the Van-Hove filling 
where the Fermi surface undergoes a Lifshitz transition.
(b) The THE conductivity ($\lambda = 0$) shows a crossover from a linear regime at small $J$ to saturation at large $J$.
The slope is independent of density $n$, while the saturating value increases with $n$. Both these behaviors can be qualitatively explained by 
analyzing $\mathcal{K}(n)$ as a function of $J$ (see appendix  \ref{app:solution-Boltzmann}).
(c) Variation of $\sigma_{xy}$  with spin-orbit coupling $\lambda$ at fixed $J/t = 10$ and $n = 0.2$.
With increasing $L_s/a$, the results rapidly converge to a finite value that is very weakly $\lambda$-dependent.
Thus there is no linear in $\lambda$ contribution in the Kubo result in agreement with the semi-classical analysis.
}
\label{fig2}
\end{figure*}

All other contributions to the Hall response involve the electric field induced perturbations to the 
distribution function determined by solving the Boltzmann equation.
We expand the distribution function to linear order in the electric field, $f = f^0 + g + \mathcal{O}(E^2)$ 
and substitute it into the Boltzmann equation with a relaxation time $\tau = \ell/v_F$ to find the equation for $g$:
\begin{eqnarray}
\left( 1 + \tau \dot{\boldsymbol\xi}^{(I)}\cdot \boldsymbol\nabla_{\boldsymbol\xi} \right) g(\boldsymbol\xi) = -\tau \;\dot{\boldsymbol\xi}^{(D)} \cdot \boldsymbol\nabla_{\boldsymbol\xi}f^0[\mathcal{E}(\boldsymbol\xi)] \label{eq:distr-fnc}
\end{eqnarray}
where $\dot{\boldsymbol\xi}^{(I)}$ and $\dot{\boldsymbol\xi}^{(D)}$ are the electric field independent and dependent parts of $\dot{\boldsymbol\xi}$ 
in eq.~\eqref{eq:eom-Gamma}.
We now take advantage of the fact that $\tau\dot{\boldsymbol\xi}^{(I)}\cdot \boldsymbol\nabla_{\boldsymbol\xi}\sim (\ell/L_s)(a/L_s) \ll 1$ 
when $\ell/L_s \ll 1$ to invert the operator on the left hand side and solve for $g(\boldsymbol\xi)$.
This is analogous to the Zener-Jones calculation \cite{Ziman2007} of the Hall conductivity 
in the weak field regime  $\omega_c\tau\ll1$. Solving the Boltzmann equation for $L_s  \ll \ell$ is technically much harder.  We will investigate 
aspects of this regime using the Kubo formalism below.

The term $g^{(1)}(\boldsymbol\xi)$ linear in $\tau$ does not contribute to the Hall conductivity and the leading order 
contribution proportional to $\tau^2$ is
\begin{equation}
g^{(2)}(\boldsymbol\xi) = \tau^2 \; \dot{\boldsymbol\xi}^{(I)}\cdot \boldsymbol\nabla_{\boldsymbol\xi} \Big( \dot{\boldsymbol\xi}^{(D)} \cdot \boldsymbol\nabla_{\boldsymbol\xi} f^0[\mathcal{E}(\boldsymbol\xi)] \Big). \label{eq:g-tau2}
\end{equation}
We emphasize that this equation involves all six curvatures along with mixed derivatives of the semi-classical eigenenergies.
Combining $g^{(2)}(\boldsymbol\xi)$ with eq.~\eqref{eq:curr} we calculate the current which is linear in $E$: 
\begin{align}
    j_x^{(2)} &= -e \tau^2\int dV_{\boldsymbol\xi} \; \dot{x}^{(I)}(\boldsymbol\xi) \; \dot{\boldsymbol\xi}^{(I)}\cdot \boldsymbol\nabla_{\boldsymbol\xi} \Big( \dot{\boldsymbol\xi}^{(D)} \cdot \boldsymbol\nabla_{\boldsymbol\xi} f^0[\mathcal{E}(\boldsymbol\xi)] \Big).
\label{eq:j-tau2}
\end{align}
\noindent
We organize the calculation of the conductivity by classifying the various terms in eq.~\eqref{eq:j-tau2} in powers of the small parameters
 $\lambda/E_F$ and $a/L_s$; see appendix \ref{app:solution-Boltzmann} for details.  Here we discuss the leading order contributions.

We first focus on the zeroth order term in $(\lambda/E_F)$. 
Without SOC, all curvatures vanish except the real-space curvature which leads to the topological Hall contribution
\begin{equation}
\sigma_{xy}^{\text{THE}} = \dfrac{e^2 \tau^2}{\hbar^3} \, n_{ \text{sk} }  \; \sum_{l=\pm}\mathcal{K}_l(\mu)\bigg|_{\lambda=0} \label{eq:sigma-T-sc}.
\end{equation}
Here $n_{ \text{sk} } = \int d^2 {\bf r} \,\hat{ {\bf m} } \cdot ( \partial_{x } \hat{\bf m} \times \partial_{y} \hat{{\bf m}})/(4\pi V)$
is the skyrmion density and 
\begin{equation}
\mathcal{K}_\pm(\mu) = \mp \hbar^4 \int \dfrac{d^2 \bm{k}}{(4\pi)}\left(\dfrac{\partial f^0_\pm}{\partial \mathcal{E}}\right) {\bf v}^T 
(\mathbb{M}^{-1}-\text{Tr}\mathbb{M}^{-1}) {\bf v} \label{eq:def-Kmu}
\end{equation}
is a Fermi surface integral that depends on the chemical potential $\mu$ (or filling $n$) and the band index.
Here ${\bf v} = \boldsymbol\nabla_{{\bf k}}\mathcal{E}(\boldsymbol\xi)/\hbar$ is the band velocity vector and 
$\mathbb{M}^{-1}_{\mu\nu} = \partial_{k_\mu,k_\nu} \mathcal{E}(\boldsymbol\xi)/\hbar^2$ is the inverse mass tensor.  The semi-classical theory illuminates the relationship between the real-space Berry curvature which is a property of the spatial evolution of the semi-classical Bloch eigenstates and the skyrmion density which is a property of the spatial evolution of the magnetization vector.  In the absence of spin-orbit coupling $\Omega_{x,y}^\pm=\mp\hat{ {\bf m} } \cdot ( \partial_{x } \hat{\bf m} \times \partial_{y} \hat{{\bf m}})/2$.
The result of eq.~\eqref{eq:sigma-T-sc} bears a striking resemblance to the canonical solution  \cite{Ziman2007} 
for the semi-classical Hall conductivity with the real space Berry curvature 
$\Omega_{x,y}$ playing the role of an external magnetic field, in agreement with the intuitive picture behind the THE.  The corresponding resistivity is independent of $\tau$ and scales as $\rho_{xy}^{\text{THE}}\sim (a/L_s)^2$.

Next we focus on terms linear in $(\lambda/E_F)$. Even though there are several terms, there is only one that is linear in $(a/L_s)$.
It originates from mixed spatial and momentum space derivatives of the semi-classical energies $\mathcal{E}(\boldsymbol\xi)$ and 
is independent of all Berry curvatures:
\begin{equation}
\delta \sigma_{xy} =-\dfrac{e^2 \tau^2}{2m} \sum_{l=\pm} \,\omega_l\, n_l \label{eq:sigma-V-sc}
\end{equation}
where $\omega_l=1/V\int d^2\bm{r} \,\,\hat{\bm{z}}\cdot (\boldsymbol{\nabla}_r\times\bm{v}_{l}(\bm{r}))$ is the 
average ``vorticity" of electrons in band $l$ with velocity $\bm{v}_l(\bm{r})$ that is linear in $\lambda$ (see appendix \ref{app:vorticity} for details) and
$n_l$ is the band-resolved density defined below eq.~\eqref{eq-ahe}.
The intuition behind this term is that real-space gradients of the magnetic texture can lead to orbital electronic motion 
akin to the dynamics induced by an external magnetic field.  
For the Rashba SOC considered here, the vorticity simplifies to $\sim \int d{\bf r} \,\, \boldsymbol\nabla_r \cdot \hat{{\bf m}}({\bf r})$.
This term has been discussed in the literature \cite{Kim2013,Akosa2018,Akosa2019,BatistaPRB2020}
as a ${\cal O}(\lambda)$ correction to the emergent magnetic field arising from skyrmions.  Here this contribution arises not from SOC corrections to the real-space Berry curvature, but instead from mixed momentum and real space derivatives of the semi-classical eigenvalues.  Like the THE the corresponding resistivity is independent of $\tau$, but instead scales as $\delta\rho_{xy}\sim (a/L_s)(\lambda/E_F)$.  We note, however, that $\delta \rho_{xy}$ vanishes identically for any periodic spin texture, like a skyrmion crystal.
More generally, for any smooth texture for which ${\bf v}({\bf r})$ has continuous first order partial derivatives,
we can use Stokes' theorem and show that the vorticity leads only to a boundary term that is
negligible in the thermodynamic limit.

All other contributions to $\rho_{xy}$, including the mixed curvature terms, are
higher order in either $(\lambda/E_F)$ (which is not relevant for experiments) or in
$(a/L_s)$ at which point the semi-classical analysis presented here is itself not applicable.
Thus we have used the semi-classical approach that treats all curvatures on equal footing to 
conclude that AHE and THE resistivities are indeed additive and the largest contribution to the Hall effect for $L_s \gg \ell$.

{\bf Kubo formula analysis}:
We next turn to the opposite limit of small skyrmions such that $a \approx k_F^{-1} \lesssim L_s \ll \ell$. 
We in fact set the mean-free path to infinity and use an exact Kubo formula to numerically calculate the Hall conductivity for a lattice model of itenerant electrons in the presence of a skyrmion crystal; see appendix \ref{app:TKNN} for details.
The starting Hamiltonian is a tight-binding generalization of eq.\eqref{eq:ham-gen} describing electrons
on a lattice with nearest neighbor hopping $t$ and Rashba SOC $\lambda$, coupled to a
background spin texture described by local moments ${\bf m}_{\bf i}$ at each lattice site ${\bf i}$.
The skyrmion crystal defines an enlarged $N_s\times N_s$ unit cell, where $N_s = L_s/a$, and results in a 
magnetic Brillouin zone with $N_\text{b} = 2N_s^2$ bands. We present here results for a triangle lattice, but as
we show in appendix \ref{app:TKNN} our results are independent of the lattice for low densities.

We use exact diagonalization to compute the energy eigenvalues and eigenfunctions of our lattice Hamiltonian and then use the 
TKNN formula  \cite{TKNN1982} to determine the Hall conductivity
in terms of the momentum-space Berry curvature {\it in the magnetic Brillouin zone}.
Note that this numerically exact procedure includes all the effects of the anomalous velocity as well as
the real-space Berry curvature arising from the skyrmions, however, unlike the semi-classical theory it is hard
to decompose the final result into AHE and THE contributions. We thus proceed as follows. We first show that 
in various limits one obtains just the AHE (in a ferromagnetic background), or just the THE (in a skyrmion crystal with
$\lambda = 0$) . Finally, we consider the full problem and gain qualitative insights into the numerical results by 
comparing them with the semi-classical results described above.

First, consider the simplest ferromagnetic case with uniform magnetization $\hat{{\bf m}}_{\bf i} = \hat{\bf z}$ (independent of ${\bf i}$).
which is just the lattice version of the continuum model analyzed in ref \cite{Xiao2010} with their $\Delta \sigma_z$ corresponding to
our $J \sigma_z$.
An AHE is seen in this case provided both $\lambda$ and $\Delta$ are non-zero.
The SOC $\lambda$ breaks the two-fold spin degeneracy of the bands everywhere 
except at the time-reversal invariant momenta (TRIM) 
where time reversal (TR) enforces a Kramers degeneracy.
A non-zero $\Delta$ destroys TR symmetry, causes band inversion, and creates Berry curvature hotspots at TRIMs which 
then lead to an enhancement of the AHE conductivity whenever the Fermi level falls near the TRIM points.

We next look at a skyrmion crystal, but set $\lambda=0$ so that there is no AHE (even though the net $M_z$ is non-zero).
The Fourier modes of the periodic texture cause scattering between momentum eigenstates and lead to band folding.
At strong coupling $J/t \gg 1$, the bands separate into two sectors with the spins aligned/anti-aligned with the local magnetic texture.
The corresponding Hall conductivity is the THE arising from non-zero skyrmion number. It shows a non-trivial dependence on 
the band filling as seen in Fig.~\ref{fig2}(a) (blue curve). Comparing this with the semi-classical THE prediction of
eq.~\eqref{eq:sigma-T-sc} (red curve)  we see that these results, though obtained in very different regimes, share some qualitative features.
Both have the same sign at each density and vanish at the van Hove filling where the Fermi surface undergoes a Lifshitz transition.

%

Next consider the $J/t$ dependence of the numerical results for the THE shown in Fig.~\ref{fig2}(b). We see a
linear regime at small $J$ crossing over to saturation at large $J$. We can gain insight into
these results by analyzing the $J$ dependence of the semi-classical THE of eq.~\eqref{eq:sigma-T-sc},
which predicts an initial slope independent of density $n$ and a saturation value that increases with $n$ (see appendix \ref{app:solution-Boltzmann})

Finally, we turn to the Kubo results for a skyrmion crystal with non-zero SOC. 
In Fig.~\ref{fig2}(c) we plot these results at strong coupling $J/t = 10$ and find that in general 
the Hall response depends on the SOC $\lambda$. 
We see that with increasing $L_s/a$, the results converge to a non-zero value 
which is very weakly dependent on $\lambda$.
The large $L_s/a$ limit allows us to make contact with the semi-classical results,
where we showed above that there is no linear in $\lambda$ contribution to the Hall conductivity.
For the parameters considered here, the AHE contribution that scales like $(\lambda/J)^2 \sim 10^{-4}$
is also negligible.

{\bf Discussion:} We have presented a complete semi-classical analysis in the weak SOC $\lambda\ll E_F$ regime for 
$a \ll \ell \ll L_s$ and demonstrated that the Hall resistivity the sum of an anomalous Hall contribution, arising from the momentum space Berry curvature
and proportional to the average out-of-plane magnetization, and a topological Hall contribution, arising from the real-space Berry curvature and 
proportional to the skyrmion density. All corrections were explicitly shown to be higher order in the small parameters.
The semi-classical results are valid for any spin texture without any assumption about its periodicity.
In the opposite limit $L_s \ll \ell = \infty$ (zero disorder) we have presented exact Kubo formula results for skyrmion crystals.

We conclude by noting effects that we have not included and questions for further study. We focussed on the intrinsic anomalous Hall effect,
arising for momentum space Berry curvature, often the dominant contribution \cite{NagaosaRMP2010} to the AHE, but did not
consider extrinsic effects such as skew and side jump scattering. We have also not analyzed non-periodic spin textures which
vary on a length scale $L_s \lesssim \ell$. Such a regime has been analyzed \cite{BouazizPRL2021} in the context of electrons scattering off a 
single skrymion with the prediction of a novel non-collinear Hall effect linear in the SOC. It would be interesting to extend our semi-classical analysis 
to this regime. 

Finally, in the semiclassical regime that we have examined in detail, with $L_s \gg \ell$, there is a novel vorticity term 
[eq.~\eqref{eq:sigma-V-sc}] that is linear in $\lambda$, but we were able to use Stokes' theorem to reduce it to a boundary term that vanishes for
periodic textures. An interesting question \cite{Addison2022arxiv} for further study is the fate of this term in the presence of singularities, such as Bloch points, 
that may act as obstructions to the use of Stokes theorem.

 \smallskip
\noindent
{\bf Acknowledgements:} 
This work was supported by NSF Materials Research Science and Engineering Center Grant DMR-2011876. 
Z.A. was also supported by the Ohio State University President’s Postdoctoral Scholars Program.
We gratefully acknowledge Roland Kawakami, Siddharth Seetharaman, Po-Kuan Wu, and Fengyuan Yang for insightful discussions. 

\bibliography{ahe-the}


\newpage
\clearpage
\onecolumngrid

\appendix

\begin{center}
{\bf APPENDICES}
\end{center}

\renewcommand{\thefigure}{A\arabic{figure}}
\setcounter{figure}{0}

\section{Semi-classical Equations of Motion with Phase Space Berry Curvatures}\label{app:sc-equations}

Semi-classical theory describes transport in terms of electron wave-packets whose width is larger than microscopic lattice scale $a$ but much smaller than mean-free path $\ell$ so that the average position ${\bf r}$ and average momentum ${\bf k}$ of the wavepacket are well-defined simultaneously.
This is in addition to their time evolution which is governed by a semi-classical Hamiltonian.
The magnetic texture presents a new length-scale related to its size $L_s$.
The construction now requires a gradient expansion which introduces an additional constraint that the width is smaller than $L_s$.
The Hamiltonian thus obtained is a function of phase-space variables $\boldsymbol\xi = (x,y,k_x, k_y)$ :
\begin{equation}
\mathcal{H}(\boldsymbol\xi) = \dfrac{\hbar^2{\bf k}^2}{2m} + {\bf d}(\boldsymbol\xi)\cdot \boldsymbol\sigma, \quad {\bf d}(\boldsymbol\xi) = a\lambda ( {\bf k} \times \hat{{\bf z}} ) - J \hat{{\bf m}}({\bf r}), \quad \mathcal{E}_\pm(\boldsymbol\xi) = \dfrac{\hbar^2{\bf k}^2}{2m} \pm |{\bf d}(\boldsymbol\xi)|
\end{equation}
and hosts six types of Berry curvatures, each corresponding to a plane in the 4D phase space
\begin{eqnarray}
\Omega^\pm_{ \alpha, \beta }(\boldsymbol\xi) &=& \pm \dfrac{1}{2} \hat{ {\bf d} }(\boldsymbol\xi) \cdot ( \partial_{\alpha } \hat{\bf d}(\boldsymbol\xi) \times \partial_{\beta } \hat{{\bf d}}(\boldsymbol\xi) )
\end{eqnarray}
\noindent
where $\pm$ label the two bands.
The curvatures introduce non-trivial Poisson bracket relations between the phase space variables that lead to corrections in the equations of motion and  the invariant measure.
Both these quantities are captured by the completely anti-symmetric matrix $[\Gamma(\boldsymbol\xi)]_{\alpha,\beta} = \Omega_{ \alpha, \beta}(\boldsymbol\xi)-[ i\sigma_y \otimes \mathds{1} ]_{\alpha,\beta}$ as defined in the main text. Here we explicitly write the expression for completeness:
\begin{eqnarray}
\hbar \begin{pmatrix} \dot{x} \\ \dot{y} \\ \dot{k_x} \\ \dot{k_y} \end{pmatrix} &=& \dfrac{1}{ \sqrt{\text{det}[ \Gamma(\boldsymbol\xi) ] } }  \left[ \begin{pmatrix}
0 & \Omega_{k_x, k_y} & -\Omega_{ y, k_y} & \Omega_{ y, k_x } \\
-\Omega_{k_x,k_y} & 0 & \Omega_{ x, k_y} & -\Omega_{ x, k_x } \\
\Omega_{y, k_y} & -\Omega_{ x, k_y} & 0 & \Omega_{ x, y } \\
-\Omega_{ y, k_x } & - \Omega_{x, k_x} & -\Omega_{ x, y } & 0
\end{pmatrix} - \begin{pmatrix}
0 & 0 & -1 & 0 \\
0 & 0 & 0 & -1 \\
1 & 0 & 0 & 0 \\
0 & 1 & 0 & 0 \\
\end{pmatrix} \right] \begin{pmatrix} \partial_x \widetilde{\mathcal{E}}(\boldsymbol{\xi}) \\ \partial_y \widetilde{\mathcal{E}}(\boldsymbol{\xi}) + e E\\ \partial_{k_x} \widetilde{\mathcal{E}}(\boldsymbol{\xi}) \\ \partial_{k_y} \widetilde{\mathcal{E}}(\boldsymbol{\xi}) \end{pmatrix} \label{eq:ap:fullEOM} \\
dV_{\boldsymbol\xi} &=& \dfrac{d^2 {\bf r} d^2 {\bf k} }{(2\pi)^2V} \sqrt{\text{det}[ \Gamma(\boldsymbol\xi) ] }
\end{eqnarray}
\noindent

There is one additional change in the equations.
The non-trivial spatial and momentum variation in the eigenfunction of the semi-classical Bloch Hamiltonian $|u(\boldsymbol\xi)\rangle$ leads to a shift in the energy $\widetilde{\mathcal{E}}(\boldsymbol\xi)=\mathcal{E}(\boldsymbol\xi)+\delta\mathcal{E}(\boldsymbol\xi)$ with
\begin{equation}
\delta\mathcal{E}(\boldsymbol\xi)  =-\sum_{i=x,y}\text{Im} \bigg[ \bigg(\partial_{r_i}\langle{u(\boldsymbol\xi)}|\bigg)(\mathcal{E}(\boldsymbol\xi)-\mathcal{H}(\boldsymbol\xi))\bigg(\partial_{k_i}|{u(\boldsymbol\xi)}\rangle\bigg)\bigg].
\end{equation}
\noindent

\noindent
We can ignore $\delta\mathcal{E}(\boldsymbol\xi)$ in our calculation because it scales as $(\lambda/E_F)(a/L_s)$ and thus leads to higher order corrections to the Hall effect not considered here.

The matrix representation in eq.~\eqref{eq:ap:fullEOM} contains all contributions of the curvatures.
In particular, the anomalous velocity can be extracted from the electric field dependent part of the velocity in $+x$ direction
\begin{equation}
\dot{x}^{(D)}(\boldsymbol\xi) = \dfrac{e}{\hbar}\dfrac{\Omega_{ k_x, k_y}(\boldsymbol\xi)}{ \sqrt{\text{det}[ \Gamma(\boldsymbol\xi) ] } }  E.
\end{equation}
The determinant factor in the denominator may seem unfamiliar but is absolutely crucial for calculating the correct intrinsic anomalous Hall response. 
There is a complete cancellation of the phase-space measure factors in the Hall current  so that the anomalous Hall conductivity only depends on the momentum-space Berry curvature:
\begin{equation}
j_x = e \int\dfrac{d^2 {\bf r} d^2 {\bf k} }{(2\pi)^2 V} \sqrt{\text{det}[ \Gamma(\boldsymbol\xi) ] } \; \dot{x}^{(D)}(\boldsymbol\xi) \; f^0(\boldsymbol\xi) = \left(\dfrac{e^2}{\hbar} \int \dfrac{d^2 {\bf r} d^2 {\bf k} }{(2\pi)^2 V}\; \Omega_{ k_x, k_y}(\boldsymbol\xi) \; f^0(\boldsymbol\xi) \right) E
\end{equation}
The quantity within brackets is $\sigma_{xy}^\text{AHE}$. Even though the expression contains only the momentum-space Berry curvature $\Omega_{ k_x, k_y}(\boldsymbol\xi)$, we must keep in mind that $\Omega_{ k_x, k_y}(\boldsymbol\xi)$ is a function of momentum and real space and the two integrals are not separable:
\begin{equation}
\sigma_{xy}^\text{AHE} = -\sum_{l=\pm} l\dfrac{e^2}{\hbar} \int \dfrac{d^2 {\bf r} \;d^2{\bf k} }{(2\pi)^2V} \;\left[ \dfrac{a^2 \lambda^2 J m_z({\bf r})}{2|{\bf d}(\boldsymbol\xi)|} \right]  \; f^0[ \mathcal{E}_l(\boldsymbol\xi) ].
\end{equation}
Since there is already an explicit $\lambda^2$ and $\lambda/E_F$ is a small parameter, we can set $\lambda=0$ in the rest of the expression to find the leading contribution.
The spatial dependence in the semi-classical eigenenergies drops out when $\lambda=0$, that is $(\mathcal{E}_l(\boldsymbol\xi; \lambda =0) = \mathcal{E}_l({\bf k}))$ and the spatial and momentum integrals become separable
\begin{align}
\sigma_{xy}^\text{AHE} & \approx -\sum_{l=\pm}l\dfrac{e^2 a^2}{2\hbar} \bigg(\dfrac{\lambda}{J}\bigg)^2\left( \int \dfrac{d^2 {\bf r}}{V} \; m_z({\bf r}) \right) \left( \int \dfrac{d^2{\bf k} }{(2\pi)^2} f^0[\mathcal{E}_l({\bf k})] \right).
\end{align}
We thus find that the intrinsic contribution only probes the net out-of-plane magnetization even for spatially varying textures.

\section{Solution to the Boltzmann Equation}\label{app:solution-Boltzmann}
Focussing on contributions that come from electric field induced perturbations to the distribution function,  we write the full distribution function in the presence of electric field as $f = f^0 + g$ where $g$ is linear order in the field.
We then use the relaxation time approximation to write the Boltzmann equation as
\begin{eqnarray}
\dot{\boldsymbol\xi}\cdot \boldsymbol\nabla_{\boldsymbol\xi} \left( f^0(\boldsymbol\xi) + g(\boldsymbol\xi) \right)= -\dfrac{g(\boldsymbol\xi)}{\tau}.
\end{eqnarray}
We write $\dot{\boldsymbol\xi} = \dot{\boldsymbol\xi}^{(I)} +\dot{\boldsymbol\xi}^{(D)}$ where $I$ and $D$ refer to electric field dependent and independent components to find an equation for $g(\boldsymbol\xi)$
\begin{equation}
\left( 1 + \tau \dot{\boldsymbol\xi}^{(I)}\cdot \boldsymbol\nabla_{\boldsymbol\xi} \right) g(\boldsymbol\xi) = -\tau \;\dot{\boldsymbol\xi}^{(D)} \cdot \boldsymbol\nabla_{\boldsymbol\xi}f^0[\mathcal{E}(\boldsymbol\xi)].
\end{equation}
The differential operator on the left has a particular scaling. With $g(\boldsymbol\xi) = g[\mathcal{E}(\boldsymbol\xi)]$, we can infer that
$\tau \dot{\boldsymbol\xi}^{(I)}\cdot \boldsymbol\nabla_{\boldsymbol\xi}\mathcal{E}\sim (\ell/L_s)(a/L_s)$.
Now since $a \ll \ell \ll L_s$, both these ratios are small and hence, we can invert the operator to find
\begin{equation}
 g(\boldsymbol\xi) = -\tau \; \left( 1 - \tau \dot{\boldsymbol\xi}^{(I)}\cdot \boldsymbol\nabla_{\boldsymbol\xi} \right)\;\dot{\boldsymbol\xi}^{(D)} \cdot \boldsymbol\nabla_{\boldsymbol\xi}f^0[\mathcal{E}(\boldsymbol\xi)] =  g^{(1)}(\boldsymbol\xi)  +   g^{(2)}(\boldsymbol\xi)
\end{equation}
where the superscripts label the order in $\tau$.
The first order term $g^{(1)}(\boldsymbol\xi)$ does not result in any Hall conductivity.
While we find that at the end of a long calculation, we can use time-reversal (TR) symmetry to understand why it vanishes. 
Onsager's reciprocity relation forces Hall conductivity to be odd under TR.
The conductivity arising from $g^{(1)}(\boldsymbol\xi)$ doesn't have this property:
 \begin{equation}
\sigma_{xy}\sim \dfrac{e}{V} \int \dfrac{d^2{\bf r} d^2 {\bf k}}{(2\pi)^2} \underbrace{\sqrt{\text{det}[\Gamma(\boldsymbol\xi)]}}_{ \text{TR even} }  \; \underbrace{\dot{x}(\boldsymbol\xi)}_{ \text{TR odd} } \; \Big( -\tau \underbrace{\dot{\boldsymbol\xi}^{(D)}}_{ \text{TR odd} } \cdot \underbrace{\boldsymbol\nabla_{\boldsymbol\xi}f^0[\mathcal{E}(\boldsymbol\xi)]  }_{ \text{TR even} } \Big).
\end{equation}
It is clearly even under time-reversal and hence must vanish. 
The contributions from $g^{(2)}(\boldsymbol\xi)$ survive this argument. 

The calculation for Hall conductivity involves combining the distribution function with velocity and the appropriate phase space volume factor.
The algebra is tedious but there are a few simplifying factors.
Products and derivatives of the curvatures can be excluded as they are all higher order in $(a/L_s)$.
There will still be many terms and hence we need to introduce a classification scheme for bookkeeping 
\begin{equation}
\prod\limits_{i=1}^m \prod\limits_{j=1}^n \partial_{ k_i } \partial_{ r_j } (\cdot) \longrightarrow (m,n) .
\end{equation}
Here $(m,n)$ labels expressions that have $m$ momentum derivatives and $n$ spatial derivatives. These numbers count both: derivatives with respect to the semi-classical energies and the implicit derivatives hidden inside the curvatures.

With these labels, the Hall conductivity is
\begin{equation}
\sigma_{xy }\; \sim \;\int \dfrac{d^2{\bf r} d^2 {\bf k}}{(2\pi)^2V}\; \underbrace{\phantom{\Big(}\text{Phase space volume} \phantom{\Big)}}_{ (0,0) +  (1,1)  } \; \times \; \underbrace{\phantom{\Big(}\text{Velocity} \phantom{\Big)}}_{ (1,0) + {\color{black}(2,1)} } \; \times \; \underbrace{\phantom{\Big(} \dot{\boldsymbol\xi}^{(I)}\cdot \boldsymbol\nabla_{\boldsymbol\xi} \phantom{\Big)} }_{ (1,1) + {\color{black}(2,2)} }  \; \times \underbrace{\Big( \dot{\boldsymbol\xi}^{(D)} \cdot \boldsymbol\nabla_{\boldsymbol\xi} f^0[\mathcal{E}(\boldsymbol\xi)] \Big)}_{ (1,0) + {\color{black}(2,1)} }.
\end{equation}
The first tuple indexes derivatives acting on the semi-classical eigenenergies, while the second tuple indexes derivatives deriving from the Berry curvatures.
Since we are focussing on contributions that involve at most one curvature,
there are only two broad categories: no curvature $(3,1)$ and one curvature $(4,2)$. The number of terms inside each category is still quite large.

We now turn to energy scaling relations and take advantage of the fact that $\lambda/E_F$ is a small parameter.
To the leading order, we find that $\partial_{k_x} \mathcal{E} \sim \lambda^0$, $\Omega_{x,y} \sim \lambda^0$, $\partial_x \mathcal{E} \sim \lambda$, , $\Omega_{x,k_y} \sim \lambda$ and $\Omega_{k_x, k_y} \sim \lambda^2$. We are now ready to calculate the contributions order by order in $\lambda/E_F$ and $a/L_s$:

\begin{itemize}
\item Zeroth order in $\lambda/E_F$

The equations of motion are quite simple since all spatial derivatives vanish. There are no $(3,1)$ terms and the only non-zero $(4,2)$ term has both spatial derivatives coming from the real-space Berry curvature.
The resulting contribution is the Topological Hall response $\sigma_{xy}^{\text{THE}}$.  Simplifying eq.~\eqref{eq:sigma-T-sc} we find
\begin{equation}
     \sigma_{xy}^{\text{THE}}=-\dfrac{e^2\tau^2}{\hbar^3}n_{\text{sk}}\dfrac{2h^2}{m}
     \begin{cases}
\dfrac{\pi\hbar n}{m}, & \text{for $0<n<\dfrac{mJ}{\pi\hbar}$} \\
  J, & \text{for $\dfrac{mJ}{\pi\hbar}<n$}
  \end{cases}
  \label{thallapp1}
\end{equation}
where $n$ is the electron density. Hence as $J/E_F$ is tuned, $\sigma_{xy}$ crosses over from a linear in $J$ regime to a saturating value that is independent of $J$ but increases with density. Here $n_\text{sk}$ is the skyrmion density
\begin{equation}
n_{\text{sk}}=\dfrac{1}{V} \int \dfrac{d^2\bm{r}}{4\pi}\,\,\hat{\bm{m}}(\bm{r})\cdot (\partial_{x} \hat{\bm{m}}(\bm{r})\times \partial_y \hat{\bm{m}}(\bm{r}))
\end{equation}
For small densities $n<\dfrac{mJ}{\pi\hbar}$ this can be written as

\begin{equation}
\sigma_{xy}^{\text{THE}}=\dfrac{ne\tau}{m}\bigg(\dfrac{\tau eB_{\text{eff}}}{m}\bigg)
\end{equation}

with $eB_{\text{eff}}=-2\pi n_{\text{sk}}$ acting like an effective magnetic field induced by the pressence of the spatially dispersive magnetic texture.

\item First order in $\lambda/E_F$

There are both $(3,1)$ and $(4,2)$ type of contributions.
We leave $(3,1)$ to the next section since it has a rather interesting origin, and focus on $(4,2)$, which has two possible origins. 

The first involves $\Omega_{x,y}$ multiplied with four momentum derivatives of energies.
We will now show that the resultant Hall conductivity is even in $\lambda$ and hence is either zeroth order (discussed above) or second order (can be ignored).
It can be checked that the semi-classical eigenenergies and the real-space curvature satisfy the relations 
\begin{equation}
\mathcal{E}({\bf r}, {\bf k}, \lambda) = \mathcal{E}({\bf r}, -{\bf k}, -\lambda); \quad \Omega_{x,y}({\bf r}, {\bf k}, \lambda) = \Omega_{x,y}({\bf r}, -{\bf k}, -\lambda).
\end{equation}
As a result, the integrand in phase space will switch $\lambda$ upon flipping the momentum ${\bf k}\rightarrow -{\bf k}$. 
The resulting Hall conductivity changes $\sigma(\lambda) \rightarrow \sigma(-\lambda)$ under the flip. 
However, since ${\bf k}$ is a dummy variable that is being integrated over, Hall conductivity must satisfy $\sigma(\lambda) = \sigma(-\lambda)$ and is hence even in $\lambda$. There cannot be any first order corrections.

The other possibility both involves mixed curvatures, which as we showed, are at least linear in $\lambda/E_F$. 
Since the overall type has to be $(4,2)$, the pre-factors that come with mixed curvature should be of the type $(3,1)$. That is, there will be an additional spatial derivative in the full expression. 
It can either come from a different mixed curvature piece or from a first order spatial derivative of the semi-classical eigenenergies. It is easy to see that both these situations lead to second order contributions.

\end{itemize}

In sum, the only linear order contribution in SOC is of type $(3,1)$. It is the subject of the next section.

\section{Hall conductivity independent of Curvatures}\label{app:vorticity}
There are many simplifications when the curvatures are absent.
We therefore find it instructive to present the full derivation, starting from the fact that the semi-classical energy is a function of both real space and momentum, $\mathcal{E}({\bf r}, {\bf k})$.
The derivation also appeals to the generality of the result and that it may apply to systems beyond the model Hamiltonian that we have considered in this paper.

With external Electric field, ${\bf E}$, the dynamics of the wave-packet is governed by the equations :
\begin{equation}
 \begin{pmatrix}
\dot{{\bf r}} \\ \dot{{\bf k}} 
\end{pmatrix} = \begin{pmatrix}
0 & \mathds{1} \\
-\mathds{1} & 0 
\end{pmatrix} \begin{pmatrix}
\boldsymbol\nabla_r \mathcal{E}/\hbar + e {\bf E}/\hbar \\ \boldsymbol\nabla_k \mathcal{E}/\hbar
\end{pmatrix}.\label{eq:eom-Gamma2}
\end{equation}
that lead to the following second-order shift in the distribution function
\begin{equation}
g^{(2)}(\boldsymbol\xi) = -\dfrac{e\tau^2}{\hbar^2} \left( \dfrac{\partial f^0}{\partial \mathcal{E}} \right) \big[ \boldsymbol\nabla_k \mathcal{E} \cdot \boldsymbol\nabla_r - \boldsymbol\nabla_r \mathcal{E} \cdot \boldsymbol\nabla_k \big]{\bf E} \cdot \boldsymbol\nabla_k \mathcal{E}
\end{equation}
and a Hall conductivity
\begin{equation}
\sigma_{\alpha\beta} = \dfrac{e^2 \tau^2}{\hbar^3} \dfrac{1}{V} \int \dfrac{d^2{\bf r} \;d^2{\bf q}}{(2\pi)^2} \; \left( \dfrac{\partial f^0}{\partial \mathcal{E}} \right)  \; (\partial_{k_\alpha} \mathcal{E}) \big[ \boldsymbol\nabla_k \mathcal{E} \cdot \boldsymbol\nabla_r - \boldsymbol\nabla_r \mathcal{E} \cdot \boldsymbol\nabla_k \big] (\partial_{k_\beta} \mathcal{E} ). \label{eq:sigma-alphabeta}
\end{equation}

\noindent
where we have suppressed the sum over the band index for brevity.

It is not obvious from the expression, as it stands, to see that the anti-symmetric response, $\sigma_{xy} - \sigma_{yx}$, is finite. Therefore, we next use integration by parts to rewrite the tensor as
\begin{equation}
\sigma_{\alpha\beta} = -\dfrac{e^2 \tau^2}{\hbar^3} \dfrac{1}{V} \int d{\bf r} \;d{\bf q} \; f^0[ \mathcal{E} ]  \; \big[ \boldsymbol\nabla_k \partial_{k_\alpha} \mathcal{E} \cdot \boldsymbol\nabla_r - \boldsymbol\nabla_r \partial_{k_\alpha} \mathcal{E} \cdot \boldsymbol\nabla_k \big] (\partial_{k_\beta} \mathcal{E} ) + \mathcal{S}_{\alpha\beta}
\end{equation}
where $\mathcal{S}$ is a symmetric tensor, $\mathcal{S}_{\alpha\beta} = \mathcal{S}_{\beta\alpha}$, and the integrand is explicitly anti-symmetric
\begin{equation}
 \big[ \boldsymbol\nabla_k \partial_{k_\alpha} \mathcal{E} \cdot \boldsymbol\nabla_r - \boldsymbol\nabla_r \partial_{k_\alpha} \mathcal{E} \cdot \boldsymbol\nabla_k \big] (\partial_{k_\beta} \mathcal{E} ) =  \boldsymbol\nabla_k \partial_{k_\alpha} \mathcal{E} \cdot \boldsymbol\nabla_r\partial_{k_\beta} \mathcal{E} - \boldsymbol\nabla_r\partial_{k_\alpha} \mathcal{E} \cdot \boldsymbol\nabla_k \partial_{k_\beta} \mathcal{E} \label{eq:symm-s2}.
\end{equation}
Thus, the net anti-symmetric part can survive.

Back to our model Hamiltonian, we see that this effect cannot be described as an anomalous or topological Hall response. 
It survives in the absence of both Berry curvatures.
As we will show now, its origin lies in vorticity of the local electronic velocity field.
We expand the integrand
\begin{equation}
\partial_{k_x}^2 \mathcal{E}  \partial_{x, k_y} \mathcal{E} + \partial_{k_x,k_y} \mathcal{E}  \partial_{y, k_y} \mathcal{E} - \partial_{x, k_x} \mathcal{E} \partial_{k_x, k_y} \mathcal{E} - \partial_{y, k_x} \mathcal{E} \partial_{k_y}^2 \mathcal{E}
\end{equation}
and use the fact that $\partial_{k_\alpha, k_\beta}\mathcal{E} =\delta_{\alpha,\beta}\; \hbar^2/m + \mathcal{O}(\lambda^2)$ to ignore the middle two terms when $\lambda/E_F$ is small. The other two terms to first order can be written as
\begin{equation}
\dfrac{\hbar^2}{m} \left( \partial_{x, k_y} \mathcal{E} - \partial_{y, k_x} \mathcal{E} \right) = \dfrac{\hbar^2}{m} \left( {\bf z} \cdot \boldsymbol\nabla_{\bf r} \times ( \boldsymbol\nabla_{\bf k} \mathcal{E} ) \right) = \dfrac{\hbar^3}{m} \left( {\bf z} \cdot \boldsymbol\nabla_{\bf r} \times \bm{v}(\bm{r}) \right) .
\end{equation}

An intuitive picture behind ordinary Hall effect is that electrons undertake cyclotron orbits under the action of the magnetic field. This results in electron velocity field forming vortices.
This contribution, on the other hand, doesn't require an external magnetic field and instead uses the underlying magnetic texture to mimic vortices.
The explicit connection to the texture is
\begin{eqnarray}
\partial_{x, k_y} \mathcal{E}_\pm - \partial_{y ,k_x} \mathcal{E}_\pm  = \mp a\lambda \left(  \partial_x m_x+ \partial_y m_y \right) = \mp a \lambda \boldsymbol\nabla \cdot \hat{m}({\bf r})
\end{eqnarray}
which has been reported elsewhere in the literature \cite{FreimuthPRB2013,Akosa2018,Akosa2019,BatistaPRB2020} as a correction to the effective magnetic field in the presence of SOC.
The resulting Hall conductivity for small densities $n<mJ/\pi\hbar$ is
\begin{equation}
\delta\sigma_{xy} =-\dfrac{n e^2 \tau}{m} \left( \dfrac{\tau \lambda a}{\hbar} \int \dfrac{d^2 {\bf r}}{V}\; \dfrac{\boldsymbol{\nabla}_r\cdot \hat{{\bf m}}({\bf r})}{2} \right).
\end{equation}

\noindent
and can be interpreted as arrising from an effective magnetic field $\sim \lambda \boldsymbol{\nabla}_r\cdot \hat{{\bf m}}(\bm{r})$.  Lastly, we note that this integral is a boundary term. Therefore unless there are singular features in the semi-classical velocity, the integral has to vanish. 
That being said the general result in eq.~\eqref{eq:sigma-alphabeta} may still be finite for systems with alternative kinetic dispersion relations $\mathcal{E}(\boldsymbol{\xi})$.

\section{Kubo Formula Calculation}\label{app:TKNN}

\begin{figure}
\centering
\includegraphics[width=0.9\textwidth]{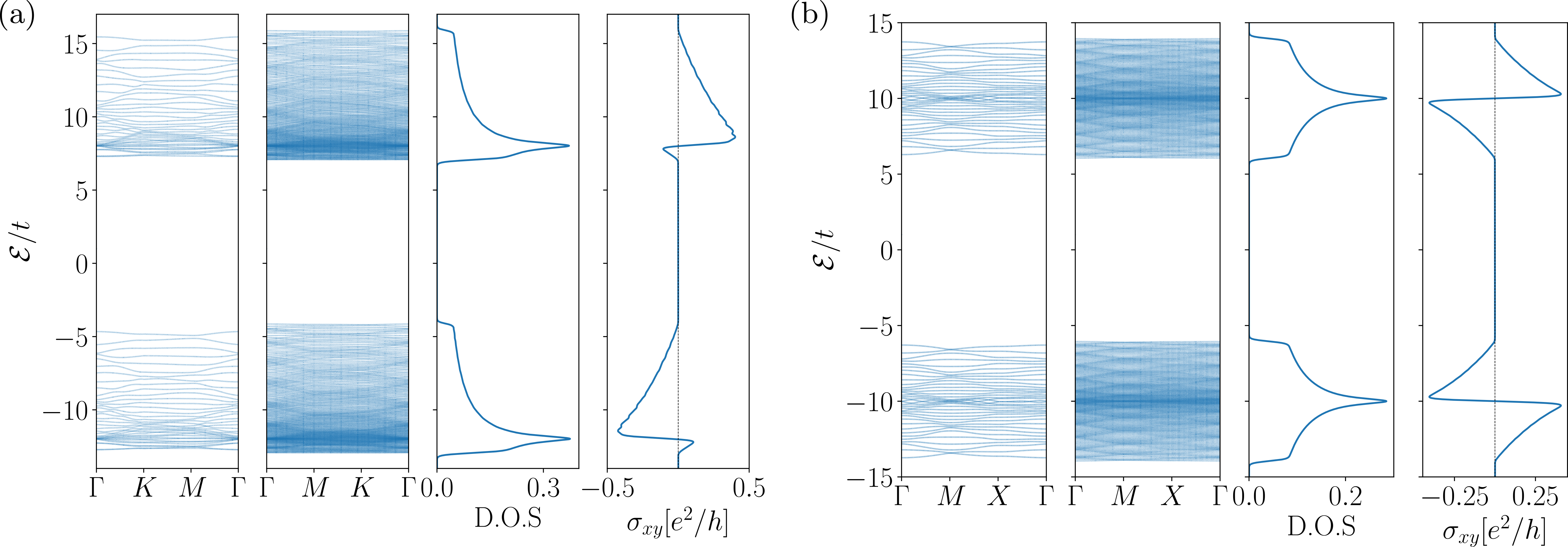}
\caption{Hall conductivity from exact diagonalization for a skyrmion crystal with (a) triangle and (b) square lattice with strong coupling $J/t=10$. The first panel shows the bands for a skyrmion unit cell with $L_s/a = 6$. The skyrmion potential causes band folding between momentum eigenstates. The number of bands increases on increasing the lattice resolution of the skyrmion, as seen in the second panel with $L_s/a = 14$. The resulting DoS and Hall conductivity are shown in the last two panels.}\label{fig:app1}
\end{figure}

While the semi-classical calculation produces intuitive results, the algebra is only controlled when $\lambda/E_F$, $(a/L_s)$, and $\ell/L_s$ are small parameters.
In this section, we will discuss the opposite limit with $a,L_s \ll \ell = \infty$.
We use a tight binding model with magnetic unit cell area $\sim (L_s/a)^2$
and calculate the Hall conductance using the TKNN Kubo formula.
The results of the calculation are exact and contain information deriving from all the types of contributions to the Hall effect.  
Guided by the semi-classical calculation here we discuss certain limiting cases.

We consider a tight-binding version of the continuum model
\begin{equation}
\mathcal{H} = -t \sum\limits_{ \langle {\bf i}, {\bf j} \rangle, \sigma } c^\dag_{ {\bf i} \sigma } c^{\phantom\dag}_{ {\bf j} \sigma } - i \lambda \sum\limits_{ \langle {\bf i}, {\bf j} \rangle, \sigma, \sigma^\prime } c^\dag_{ {\bf i} \sigma } \left[ {\bf r}_{ij} \times {\bf z}\cdot\boldsymbol\sigma \right]_{\sigma\sigma^\prime} c^{\phantom\dag}_{ {\bf j} \sigma^\prime } - J \sum\limits_{ {\bf i}, \sigma, \sigma^\prime } c^\dag_{ {\bf i} \sigma } \left[ \hat{{\bf m}}_{i}\cdot\boldsymbol\sigma \right]_{\sigma\sigma^\prime} c^{\phantom\dag}_{ {\bf i} \sigma^\prime }
\end{equation}
where the vector field $\hat{{\bf m}}_{\bf i}$ models a discrete version of a skyrmion
\begin{equation}
{\bf m}_{ {\bf i} } = \begin{pmatrix}
\sin\left( 2\pi {\bf i}\cdot {\bf a}_1 \right) \\
\sin\left( 2\pi {\bf i}\cdot {\bf a}_2 \right) \\
\cos\left( 2\pi {\bf i}\cdot {\bf a}_1 \right) + \cos\left( 2\pi {\bf i}\cdot {\bf a}_2 \right) + 1 
\end{pmatrix}, \quad \hat{{\bf m}}_{\bf i} = \dfrac{{\bf m}_{ {\bf i} }}{\sqrt{{\bf m}_{ {\bf i} }\cdot {\bf m}_{ {\bf i} }}}
\end{equation}
with winding number $+1$. Here ${\bf a}_1$ and ${\bf a}_2$ are the lattice vectors for the skyrmion lattice and ${\bf i}$ labels a position inside the skyrmion unit cell.
The corresponding magnetic Brillouin Zone (MBZ) is spanned by vectors ${\bf b}_1$ and ${\bf b}_2$ that satisfy ${\bf a}_i \cdot {\bf b}_j = 2\pi \delta_{i,j}$. These vectors permit a momentum representation
\begin{equation}
c^{\phantom\dag}_{ {\bf k} \sigma} = \dfrac{1}{\sqrt{N_{uc}}} \sum\limits_{ {\bf k} \in \text{MBZ} } e^{ i {\bf k}\cdot {\bf i} } c^{\phantom\dag}_{ {\bf i} \sigma}
\end{equation}
with ${\bf k}$ taken from a $N_k \times N_k$ BZ mesh.
The Bloch Hamiltonian which is then diagonalized to find the energies and wavefunctions
\begin{equation}
\mathcal{H}({\bf k}) | u_{ n, {\bf k} } \rangle =\epsilon_n({\bf k})  | u_{ n, {\bf k} } \rangle.
\end{equation}
The wave-functions lead to the Berry curvature
\begin{equation}
\Omega_n({\bf k}) = - 2 \text{Im} \langle \partial_{k_x} u_{ n, {\bf k} } | \partial_{k_y} u_{ n, {\bf k} } \rangle
\end{equation}
that is then combined with the TKNN formula to calculate the Hall conductivity 
\begin{equation}
\sigma_{xy} = -\dfrac{e^2}{\hbar} \dfrac{1}{V} \int\limits_\text{MBZ} \dfrac{d^2 {\bf k}}{(2\pi)^2} \sum\limits_{ n=1}^{N_b} \Omega_n ({\bf k})\; \Theta(\mu - \epsilon_n({\bf k}) ).
\end{equation}
Finally, we replace the integral by a discrete sum
\begin{equation}
\dfrac{1}{V} \int\limits_\text{MBZ} \dfrac{d^2 {\bf k}}{(2\pi)^2}  \longrightarrow \dfrac{1}{ \mathcal{V} N_{k}^2} \sum\limits_{ {\bf k} \in \text{MBZ} }.
\end{equation}
where $\mathcal{V} = | {\bf a}_1 \times {\bf a}_2| $ is the area of the unit cell.
We chose the normalization so that the density of electrons per unit cell
\begin{equation}
n =\dfrac{1}{ N_s^2 N_{k}^2} \sum\limits_{ n,{\bf k} \in \text{MBZ} }\Theta(\mu - \epsilon_n({\bf k}) ).
\end{equation}
goes from 0 (empty) to 2 (filled) as the chemical potential $\mu$ is varied across the spectrum (see Fig.~\ref{fig:app1}).

The resolution of the skyrmion within the unit cell is controlled by $N_s = L_s/a$.
Larger $N_s$ lead to a better real-space mesh but also give rise to a larger Bloch Hamiltonian with $2 N_s^2$ bands.
The bottleneck in our code is the matrix diagonalization step whose complexity is $\mathcal{O}(n^3)$ where $n = N_s^2$ is the size of the matrix. 
Since this step has to be repeated $N_k^2$ times, the overall complexity is $\mathcal{O}( N_k^2 N_s^6)$ and hence the continuum limit $(N_s\rightarrow \infty$ is more difficult than thermodynamic limit $N_k \rightarrow \infty$.
We find that $N_k \rightarrow \infty$ and $N_s\rightarrow \infty$ limits can be different, especially for the vorticity correction term which is sensitive to $N_s$.

\end{document}